\documentclass[11pt,a4paper]{article}
\pdfoutput=1
\usepackage{jheppub}
\usepackage{gensymb}
\usepackage{placeins}
\usepackage{multirow}
\usepackage{dcolumn}
\usepackage{subfigure}
\usepackage{booktabs} % For \toprule, \midrule, and \bottomrule
\usepackage{amssymb,amsmath}
\usepackage{graphicx}
\usepackage{color}
\usepackage{cancel}
\usepackage{caption}
\renewcommand{\figurename}{Fig.}

\begin{document}
\begin{titlepage}
\pagestyle{empty}

\vspace*{0.2in}
\begin{center}
{\Large \bf Gravitational wave emission from metastable current-carrying strings in $E_6$ \\
\vspace{1cm}}
{\bf Adeela Afzal$^{a,}$\footnote{E-mail: adeelaafzal555@gmail.com}, Qaisar Shafi$^{b,}$\footnote{E-mail: qshafi@udel.edu} and Amit Tiwari$^{b,}$\footnote{E-mail: amitiit@udel.edu}}
\vspace{0.5cm}

{\small \it
$^a$Department of Physics, Quaid-i-Azam University, Islamabad, 45320, Pakistan\\
$^b$Bartol Research Institute, Department of Physics and Astronomy,
University of Delaware, Newark, DE 19716, USA
}

\end{center}

\vspace{0.5cm}
\begin{abstract}

We discuss $E_6$ based extensions of the Standard Model (SM) containing two varieties of superheavy metastable cosmic strings (CSs) that respectively have neutral and electrically charged current carriers. We employ an extended version of the velocity-dependent one-scale (VOS) model, recently discussed by some authors, to estimate the gravitational wave (GW) spectrum emitted by metastable strings with a dimensionless string tension $G \mu \approx 10^{-6}$ that carry a right-handed neutrino (RHN) current. We find that with a low to moderate amount of current, the spectrum is compatible with the LIGO O3 run and also consistent at the 1$\sigma$ level with the recent PTA signals.

\end{abstract}
\end{titlepage}

\section{Introduction}
\label{sec:intro}
It has been known for some time that topologically stable strings are predicted in $SO(10)$ grand unified models if the symmetry breaking to $SU(3)_c \times U(1)_\text{em}$ is implemented using only tensor representations \cite{Kibble:1982ae}. The detection of gravitational waves (GWs) emitted by such cosmic strings (CSs) and a measurement of the dimensionless string tension parameter should provide crucial information about the symmetry breaking scale associated with the appearance of these strings. Typically, this latter scale lies well beyond the reach of high energy colliders. A number of Pulsar Timing Array (PTA) experiments \cite{NANOGrav:2023gor,NANOGrav:2023hvm,EPTA:2023fyk,Reardon:2023gzh,Xu:2023wog} have recently reported evidence for the presence of a stochastic gravitational wave background (SGWB) in the nanohertz frequency range. This is potentially exciting news for the scenario based on CSs that emit GWs in a wide frequency range, including the nanohertz region relevant for the above experiments. It appears to be generally accepted that an adequate explanation of the latest PTA data in terms of CSs requires the dimensionless string tension parameter $G\mu$ to be of order $10^{-7}$ in order to also be consistent with the LIGO O3 measurements in the 20–76.6 Hz band \cite{KAGRA:2021kbb}. Moreover, these CSs must be metastable \cite{Buchmuller:2023aus} (or quasi-stable \cite{Lazarides:2023ksx}), and decay within a certain time frame. 
A number of realistic models, mostly based on $SO(10)$ (more precisely $Spin(10)$) and its various subgroups, have been put forward \cite{Buchmuller:2020lbh,Buchmuller:2021mbb,Afzal:2022vjx,Ahmed:2022rwy,Saad:2022mzu,DiBari:2023mwu,Buchmuller:2023aus,Antusch:2023zjk,Lazarides:2023rqf,Ahmed:2023rky,Maji:2023fhv,Ahmed:2023pjl,Afzal:2023cyp,Fu:2023mdu,Lazarides:2023ksx,Lazarides:2023bjd} to implement this explanation for the SGWB. For a recent discussion of topological structures in GUTs, see Refs.~\cite{Lazarides:2019xai,Chakrabortty:2019fov,Lazarides:2023iim,Dunsky:2021tih}. For an explanation of PTA observation in terms of domain walls, see Ref.~\cite{King:2023ayw}. 

Metastable CSs with $G\mu\simeq 10^{-6}$ provide a somewhat better fit to the PTA data, but they are in conflict with the LIGO O3 run unless the string network undergoes a period of inflation during its evolution \cite{Lazarides:2023rqf}. In this paper, we explore a different scenario where the CSs carry a certain amount of current associated with the right-handed neutrino (RHN) zero modes. We find that in this case $G\mu\simeq 10^{-6}$ is compatible with the LIGO O3 run and also provides a fit at 1$\sigma$ level to the recent NANOGrav signal.

Witten showed that superconducting strings appear in realistic extensions of the Standard Model (SM) \cite{Witten:1984eb}. In particular, superconducting strings readily appear in $E_6$ grand unification \cite{Witten:1984eb,Widrow:1988rc,Hebbar:2017fit,Ma:1995xk,Ma:1996fz,Keith:1997zb,King:2005jy,King:2005my,Athron:2009ue,Athron:2009bs} as well as other extensions of the SM \cite{Lazarides:1984zq,Lazarides:1986di,Lazarides:1987rq}, and they radiate both electromagnetic and gravitational radiation. In this paper, we discuss $E_6$ inspired extensions of the SM that yield two varieties of metastable CSs that respectively carry neutral and electrically charged particles. Focusing on the neutral case, we find that the GW spectrum emitted by CSs with $G\mu \approx 10^{-6}$ and accompanied by a low to moderate amount of neutral current is compatible with the NANOGrav and LIGO O3 data. For a discussion of how the CSs may capture RHNs, see Ref.~\cite{Chavez:2002sm}.

The paper is organized as follows: In Section~\ref{E6}, we discuss three distinct symmetry breaking chains in an $E_6$ model that yield the two varieties of metastable strings that carry neutral and electrically charged currents. In Section~\ref{sec:CSs}, we display the GW spectrum emitted by the metastable strings carrying neutral current over a wide frequency range, taking into account the presence of the RHN zero modes. We employ the velocity-dependent one-scale (VOS) model, recently discussed in \cite{Rybak:2022sbo,Auclair:2022ylu}, to incorporate this feature of the metastable string in the estimation of the emitted GW spectrum. [Note that previous studies incorporating a neutral current only worked with stable strings \cite{Martins:2020jbq, Martins:2021cid}.] Our conclusions are summarized in Section~\ref{Con}.

\section{Metastable current-carrying strings in \texorpdfstring{\(E_6\)}{E6}}

\label{E6}
%%%%%%%%%%%%%%%%%%%%%%%%%%%%%%%

\begin{figure}[h!]
\centering
\includegraphics[width=0.55\textwidth]{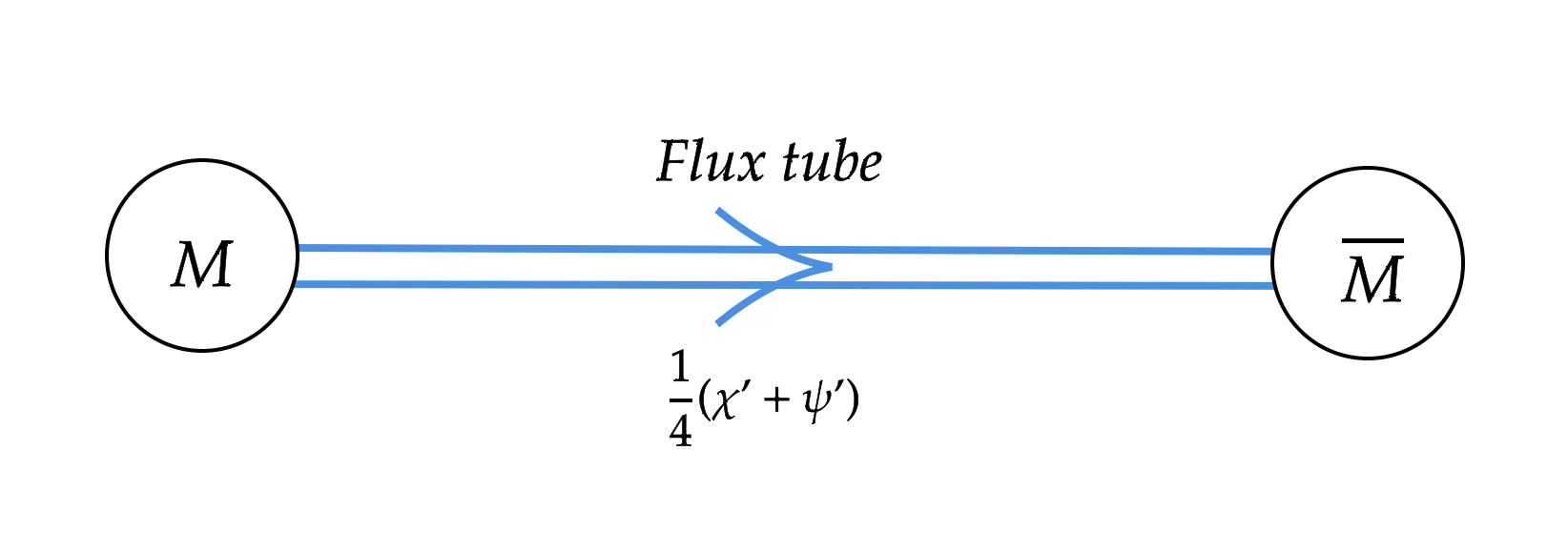}
\caption{Monopole-antimonopole dumbbell configuration from the symmetry breaking pattern summarized in Eq.~(\ref{seqe6su5}). The monopole-antimonopole pair arises from the breaking of $E_6$ to $SO(10)\times U(1)_{\psi}$.}
\label{fig:mmbarconfined}
\end{figure}

The $15$ chiral matter fields per family of the SM reside in the $27$ dimensional fundamental representation of $E_6$, and the quantum numbers of the $27$ fields under $G_\text{SM}$ are listed in Table~\ref{tab:fields}. 
\begin{table}[h!]
\centering
\scalebox{0.9}{
\begin{tabular}{|c|c|c|c|c|c|}

\hline
\hline
{Matter fields} & {Representations under $G_{\rm SM}$} & {$2\sqrt{10}\,Q_{\chi}$} & {$2\sqrt{6}\,Q_{\psi}$} & {$2\sqrt{6}\,Q_{\chi'}$} & {$2\sqrt{10}\,Q_{\psi'}$} \\

{} & {} & {} & {} & {} & {}\\

\hline

{$q_i$} & {$({\bf 3, 2}, 1/6)$} & $-1$ & $1$ & $-1$ & $1$\\

\hline

{$u_i^c$} & {$({\bf \bar 3, 1},-2/3)$} & $-1$ & $1$ & $-1$ & $1$\\

\hline

{$d_i^c$} & {$({\bf \bar 3, 1},1/3)$} & $3$ & $1$ & $2$ & $2$\\

\hline

{$l_i$} & {$({\bf 1, 2}, -1/2)$} & $3$ & $1$ & $2$ & $2$\\

\hline

{$\nu_i^c$} & {$({\bf 1, 1}, 0)$} & $-5$ & $1$ & $-4$ & $0$\\

\hline

{$e_i^c$} & {$({\bf 1, 1}, 1)$} & $-1$ & $1$ & $-1$ & $1$\\

\hline

{$H_{ui}$} & {$({\bf 1, 2},1/2)$} & $2$ & $-2$ & $2$ & $-2$\\

\hline

{$H_{di}$} & {$({\bf 1, 2},-1/2)$} & $-2$ & $-2$ & $-1$ & $-3$\\

\hline

{$D_i$} & {$({\bf 3, 1},-1/3)$} & $2$ & $-2$ & $2$ & $-2$\\

\hline

{$D_i^c$} & {$({\bf \bar 3, 1},1/3)$} & $-2$ & $-2$ & $-1$ & $-3$\\

\hline

{$N_i$} & {$({\bf 1, 1},0)$} & $0$ & $4$ & $-1$ & $5$\\

\hline
\hline
\end{tabular}
}

\caption{Transformation properties of the matter fields under the SM gauge group \( G_\text{SM} \), and their charges under the local symmetries \( U(1)_{\chi} \), \( U(1)_{\psi} \), \( U(1)_{\chi'} \) and \( U(1)_{\psi'} \). Here $Q_{\chi'}=1/4(\sqrt{15}\, Q_{\chi}-Q_{\psi})$ and $Q_{\psi'}=1/4(Q_{\chi}+\sqrt{15} \,Q_{\psi})$, where $Q_{\chi}$ and $Q_{\psi}$ are the normalized GUT generators of $U(1)_\chi$ and $U(1)_\psi$ respectively. Family indices are denoted by the subscript $i\;(=1,2,3)$.}
\label{tab:fields}
\end{table}
In addition to the well-known RHN $\nu^c_i$, we also have $SO(10)$ $10$-plets ($H_{ui},\;H_{di},\;D_{i},\;D_{i}^c$) and $SO(10)$ singlets $N_i$, where $i\;(=1,2,3)$ denotes the family index. For our first $E_6$ model containing metastable current-carrying CSs, consider the symmetry breaking chain (for related discussion including $E_6$ breaking via $SU(3)_c\times SU(3)_L\times SU(3)_R$, see Ref.~\cite{Lazarides:2019xai}): 
\begin{equation}
\begin{aligned}
    E_6 &\longrightarrow SO(10) \times U(1)_\psi\\ 
    &\longrightarrow SU(5) \times U(1)_{\chi} \times U(1)_{\psi}\\
    &\longrightarrow G_\text{SM} \times U(1)_{\chi} \times U(1)_{\psi}\\
    &\longrightarrow G_\text{SM} \times U(1)_{\psi'}\\
    &\longrightarrow G_\text{SM}.
\end{aligned}
\label{seqe6su5}
\end{equation}
In Eq.~(\ref{seqe6su5}), following Ref.~\cite{Lazarides:2019xai}, the first breaking produces an $E_6$ monopole that carries a magnetic flux $(\psi + \chi) /4 = (\psi' + \chi') /4$, where the relationship between these charges are summarized in Table \ref{tab:fields}. The second symmetry breaking produces an $SO(10)$ monopole carrying $U(1)_{\chi}$ and $SU(5)$ magnetic charges \cite{Lazarides:2023iim}. The third breaking yields the symmetry group $G_{SM} \times U(1)_{\chi} \times U(1)_{\psi}$ as well as the unconfined GUT monopole. At this stage we invoke inflation to inflate away these three varieties of monopoles, two of which are, in principle, confined and play a role in forming the metastable string networks. To implement the fourth breaking in Eq.~(\ref{seqe6su5}), we employ a complex Higgs scalar with the quantum numbers of the RHNs $\nu^c_i$, which breaks the symmetry $U(1)_{\chi} \times U(1)_{\psi}$ to $U(1)_{\psi'}$. This yields the first set of current-carrying metastable strings with the $\chi'$ flux now confined in a tube which we depict in Fig.~\ref{fig:mmbarconfined} in a dumbbell configuration.
%%%%%%%%%%%%%%%%%%%%%%%
%%%%%%%%%%%%%%%%%%%%%%%
The electrically neutral current is associated with the RHN zero modes on these strings.
The monopole-antimonopole pair in Fig.~\ref{fig:mmbarconfined} comes from the $E_6$ breaking to $SO(10) \times U(1)_{\psi}$ in Eq.~(\ref{seqe6su5}). The $\psi'$ flux carried by this pair is also confined following the breaking of $U(1)_{\psi'}$ by a Higgs scalar $N$ with the same quantum numbers as $N_i$. This latter breaking of $U(1)_{\psi'}$ provides a second set of metastable strings which happen to be superconducting because of the zero modes associated with the new electrically charged matter fields shown in Table~\ref{tab:fields} \cite{Witten:1984eb}.

 We will assume, without providing any details, that in contrast to the first set of metastable strings carrying RHN zero modes, the electrically superconducting strings and their associated monopoles and antimonopoles do not play any role in the generation of GWs.
 One possible way to implement this would be to assume that the latter set of topological structures is inflated away. For a related discussion of how this sort of scenario can be realized, see Ref.~\cite{Lazarides:2023rqf}. Of course, the electrically superconducting strings are effectively stable if the symmetry breaking scale of $U(1)_{\psi'}$ is adequately lower than the corresponding monopole mass.

Finally, since we are interested in GUT scale metastable CSs, we also assume that the symmetry breaking scales in Eq.~(\ref{seqe6su5}) are more or less comparable in magnitude to $M_{GUT} \sim 10^{16}$ GeV, the scale of $SU(5)$ symmetry breaking.

\begin{figure}[h!]
\centering
\includegraphics[width=0.6\textwidth]{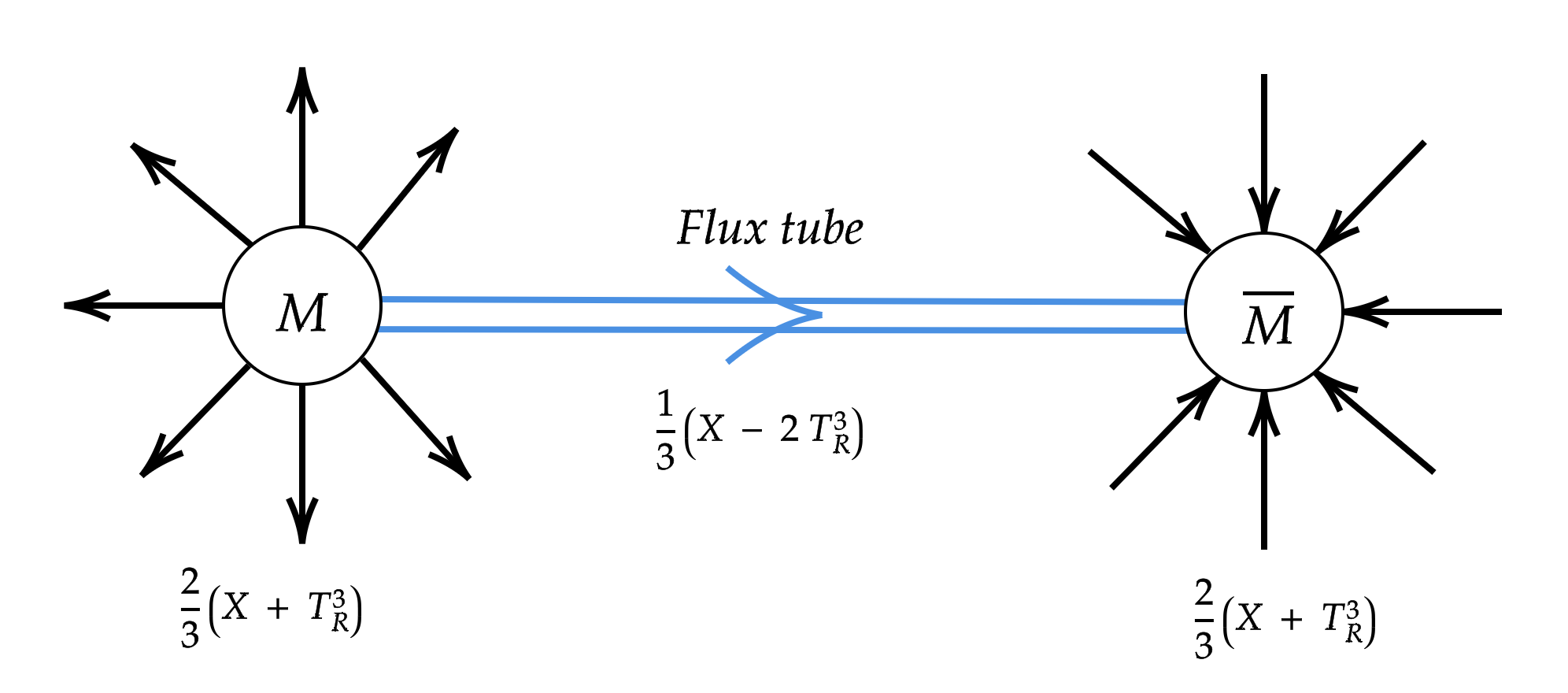}
\caption{Monopole-antimonopole dumbbell configuration from the symmetry breaking $SU(4)_c \times SU(2)_L \times U(1)_R \rightarrow SU(3)_c \times U(1)_{B-L} \times SU(2)_L \times U(1)_R \rightarrow G_{SM}$. As described in the text, the first breaking produces an $SU(4)_c$ monopole that is confined by the flux tube from the breaking of $U(1)_{B-L} \times U(1)_R \rightarrow U(1)_Y$. After electroweak breaking the confined $SU(4)_c$ monopole carries some color magnetic flux as well as Coulomb flux corresponding to a Dirac magnetic charge of $8 \pi/ 3e.$}
\label{fig:mmbar1}
\end{figure}

For our second $E_6$ model containing metastable CSs, consider the breaking chain:
\begin{equation}
    \begin{aligned}
     E_6 &\longrightarrow SO(10) \\
     &\longrightarrow SU(4)_c \times SU(2)_L \times U(1)_R \\
     &\longrightarrow SU(3)_c \times U(1)_{B-L} \times SU(2)_L \times U(1)_R \\
     &\longrightarrow G_\text{SM}. 
     \end{aligned}
     \label{eq:sequence}
\end{equation}
The first breaking by the Higgs field N leaves unbroken the $Z_5$ subgroup in $U(1)_\psi$, but no strings appear since this $Z_5$ is also contained in $SU(5)$. The second breaking yields the unconfined GUT monopole, and the third breaking produces an $SU(4)_c$ monopole \cite{Lazarides:2019xai}, whose flux is confined in the fourth step in Eq.~(\ref{eq:sequence}).
Following the discussion in Ref.~\cite{Afzal:2023cyp}, this latter system forms the superheavy metastable string structure with RHN zero modes, which produces the GW spectrum corresponding to $G\mu \approx 10^{-6}$. In Fig.~\ref{fig:mmbar1}, we show the monopole-antimonopole configuration before electroweak breaking. We have defined the generators $T_R^3 = \text{diag} (1, -1)$ and $X = (B-L) + 2\, T_c^8 /3$, where $B-L$ is an $SU(4)_c$ generator and $T_c^8 = \text{diag} (1,1,-2)$ is the color hypercharge. The monopole-antimonopole pair in this dumbbell configuration also carries unconfined Coulomb magnetic flux corresponding to a Dirac magnetic charge of $8\,\pi/3\,e$.

%%%%%%%%%%%%%%%%%%%%%
%%%%%%%%%%%%%%%%%%%%%%

\begin{figure}[h!]
\centering
\includegraphics[width=0.6\textwidth]{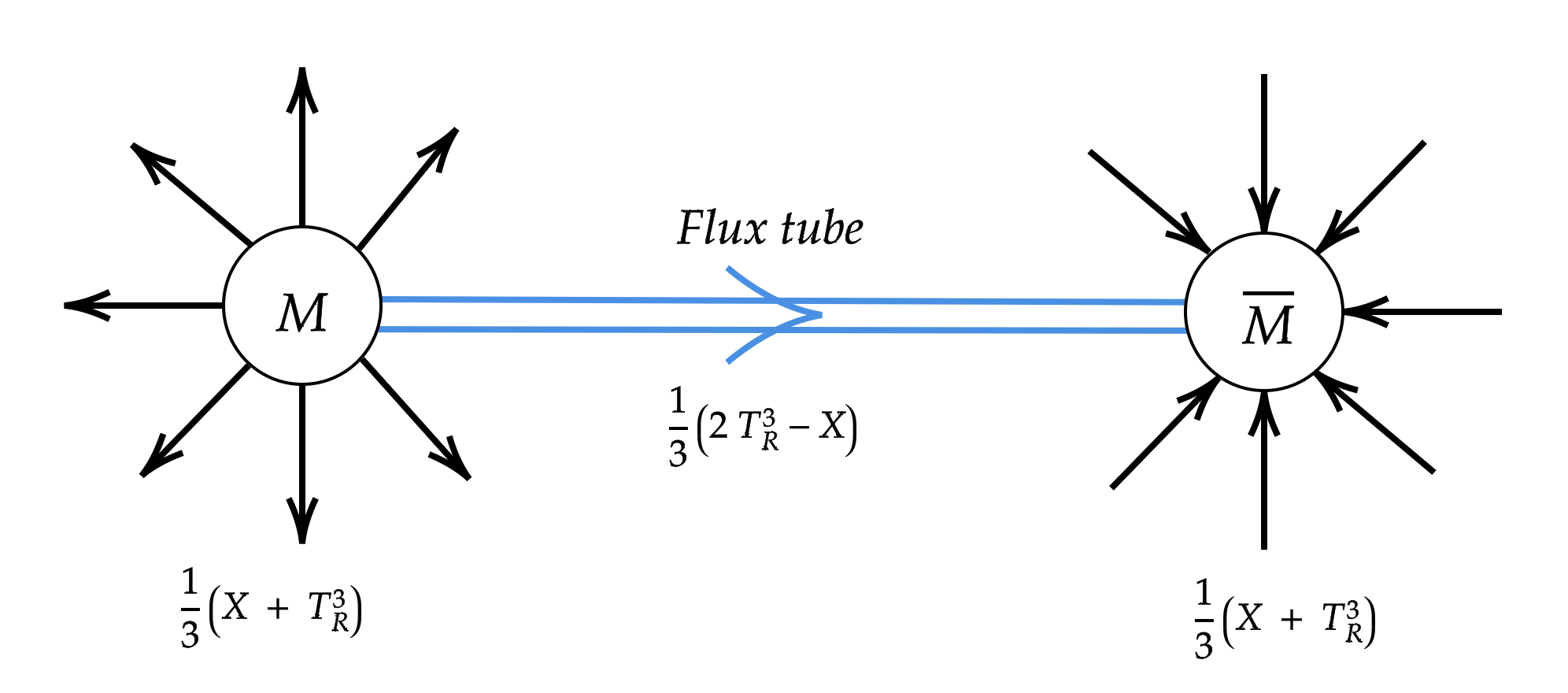}
\caption{Monopole-antimonopole dumbbell configuration from the symmetry breaking $SU(3)_c \times SU(2)_L \times SU(2)_R \times U(1)_{B-L} \rightarrow SU(3)_c \times SU(2)_L \times U(1)_R \times U(1)_{B-L} \rightarrow G_{SM}$. As described in the text, the first breaking produces an $SU(2)_R$ monopole that is confined by the flux tube from the breaking of $U(1)_{B-L} \times U(1)_R \rightarrow U(1)_Y$.  After electroweak breaking, the confined $SU(2)_R$ monopole carries a Coulomb flux corresponding to a Dirac magnetic charge of $4 \pi/3e$. The monopole also carries color magnetic charge.
}
\label{fig:mmbar2}
\end{figure}

For our third model, consider the following symmetry breaking chain:
\begin{equation}
    \begin{aligned}
     E_6 &\longrightarrow SO(10) \\
     &\longrightarrow SU(3)_c \times SU(2)_L \times SU(2)_R \times U(1)_{B-L} \\
     &\longrightarrow SU(3)_c \times SU(2)_L \times U(1)_{R}  \times U(1)_{B-L} \\
     &\longrightarrow G_\text{SM}. 
     \end{aligned}
     \label{eq:lastsequence}
\end{equation}
The spontaneous breaking of $SO(10)$ to the left-right symmetric group $H= SU(3)_c \times SU(2)_L \times SU(2)_R \times U(1)_{B-L}$ produces the unconfined GUT monopole, which can be explicitly checked by computing the first homotopy group of $H$ and taking into account that inside $SO(10)$, $H$ is more precisely written as $SU(3)_c\times SU(2)_L\times SU(2)_R\times U(1)_{B-L}/(Z_3\times Z_2)$. The GUT magnetic monopole, as expected, carries one unit of Dirac magnetic charge ($2\, \pi /e$) as well as color magnetic charge. In order to produce a metastable string network, we next break $SU(2)_R$ to $U(1)_R$, which yields a monopole, followed by the breaking of $U(1)_R \times U(1)_{B-L}$ to $U(1)_Y$, which produces the string that confines this $SU(2)_R$ monopole. Again, the RHNs $\nu^c_i$ acquire masses from the last breaking and appear as zero modes on the string. In Fig.~\ref{fig:mmbar2}, we show the monopole-antimonopole dumbbell configuration before electroweak breaking. Note that similar to Fig.~\ref{fig:mmbar1}, the monopole-antimonopole pair also carries unconfined Coulomb magnetic flux corresponding to a Dirac magnetic charge of $4\,\pi/3\,e$.

%%%%%%%%%%%%%%%%%%%%%%%

In summary, we have described three distinct $E_6$ symmetry breaking models that yield the current-carrying metastable string network we are interested in. Models 2 and 3 arise from the $SO(10)$ subgroup of $E_6$, and therefore the $SO(10)$ singlet scalar field $N$ does not play any role here. In other words, the new fermions in the $10-$plet of $SO(10)$, which acquire masses from the vacuum expectation value (VEV) of $N$, do not couple to this string and therefore do not provide electrically charged currents on it. The neutral current on the strings in models $2$ and $3$ is carried by
the right-handed neutrino fields $\nu_i^c$.
The symmetry breaking in model 1 produces, in addition to the metastable network we are interested in, an electrically superconducting string from the breaking of $U(1)_{\psi^{'}}$. If this breaking scale is sufficiently below the associated monopole mass, this string will effectively be topologically stable. The astrophysical consequences of electrically superconducting strings have been extensively studied in the literature (see Refs. \cite{HILL198717,Abe:2022rrh,Cyr:2023iwu,Theriault:2021mrq,Miyamoto:2012ck}).
%%%%%%%%%%%%%%%%%%%%%%%%%%

\section{Gravitational waves from metastable current-carrying strings}
\label{sec:CSs}

The superheavy metastable strings we are interested in carry the RHN zero mode chiral current, with the GWs as their primary decay channel. Our analysis here is based on the VOS model for the evolution of CSs, and for a detailed discussion we refer the reader to Refs.~\cite{Auclair:2019wcv,Sousa:2020sxs,Buchmuller:2021mbb}. We assume that the major contribution to GWs arise from loops (L) characterized by the loop size parameter, $\alpha=0.1$ with cusps, or loops plus segments (LS), that correspond to monopoles and antimonopoles carrying some unconfined or no unconfined Coulomb magnetic flux respectively. We have summed over $10^5$ harmonic modes and also taken into account the effective degrees of freedom in computing the GW spectra. We also assume that the zero mode chiral current associated with the metastable CSs, quantified by $Y\;(0\leq Y \leq 1)$, the averaged current strength, is present from their formation until decay. We follow the treatment of Ref.~\cite{Rybak:2022sbo} and implement it here for the metastable CSs. A slightly more conservative treatment of current-carrying CS is available in \cite{Auclair:2022ylu} where the evolution of the power emission factor $\Gamma$ of GWs is not discussed but is later explored in \cite{Rybak:2022sbo}. 

\begin{figure}[h!]
\centering
\includegraphics[width=0.70\textwidth]{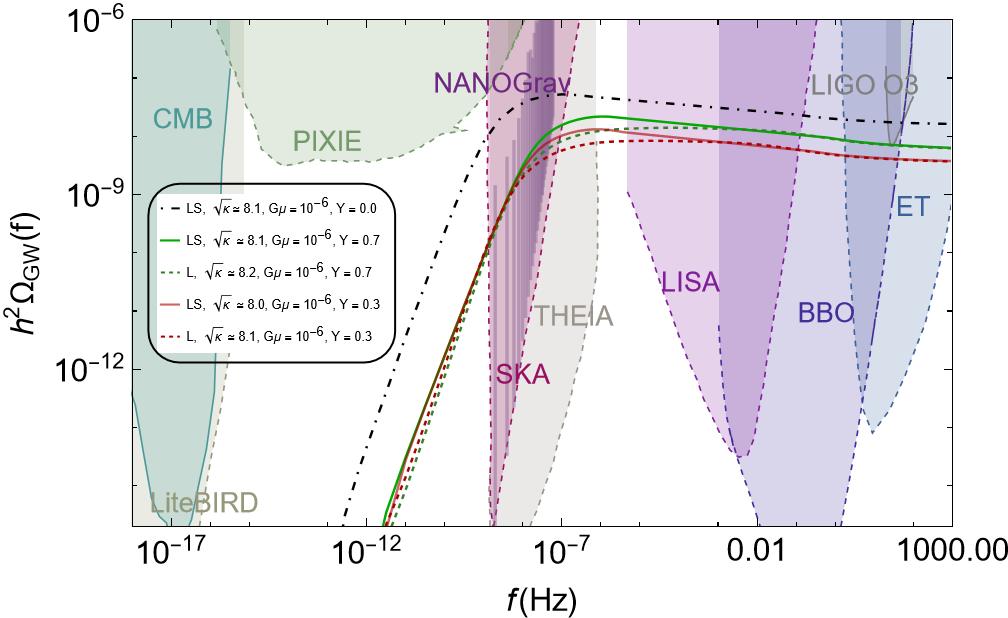}
\caption{GW spectra of metastable current-carrying CSs for $G\mu\simeq 10^{-6}$ and $Y=0.3$ ($0.7$), with $\sqrt{\kappa}\simeq 8.0$ ($8.1$) for loops plus segments (LS), solid lines and $\sqrt{\kappa}\simeq 8.1$ ($8.2$)  for loops (L) only, dashed lines. This is compatible with the LIGO O3 measurements and also explains the NANOGrav excess at 1$\sigma$ level. See also Fig.~\ref{fig:AGamma}. For comparison, we have included the standard case of $Y=0$ with LS, a dot-dashed line, which is in conflict with the LIGO O3 run. The colored shaded regions indicate the sensitivity curves of the present (solid boundaries) LIGO O3 \cite{KAGRA:2021kbb}, CMB \cite{Lasky:2015lej} and future (dashed boundaries) LiteBIRD \cite{LiteBIRD:2022cnt}, PIXIE \cite{A_Kogut_2011}, SKA \cite{Smits:2008cf}, THEIA \cite{Garcia-Bellido:2021zgu}, LISA \cite{amaroseoane2017laser}, ET \cite{Punturo:2010zz}, BBO \cite{Corbin:2005ny} experiments.}
\label{fig:GW}
\end{figure}

\begin{figure}[h!]
\centering
\includegraphics[width=0.70\textwidth]{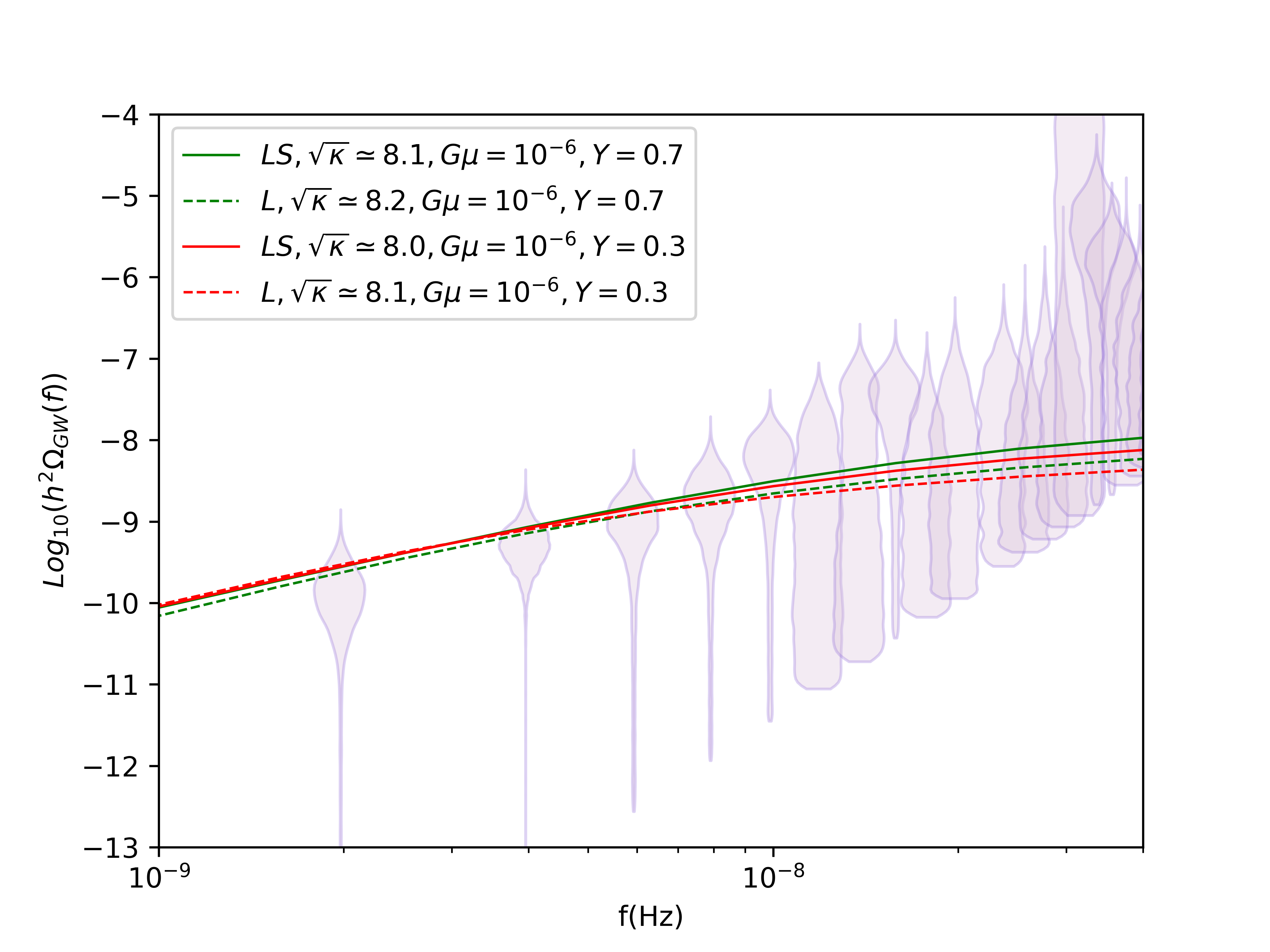}
\caption{Same as Fig.~\ref{fig:GW} but in the NANOGrav sensitivity band. The floral-violet violins are the NANOGrav frequency bins \cite{NANOGrav:2023gor}.}
\label{fig:freqbins}
\end{figure}

\begin{figure}[h]
\centering
\includegraphics[width=0.70\textwidth]{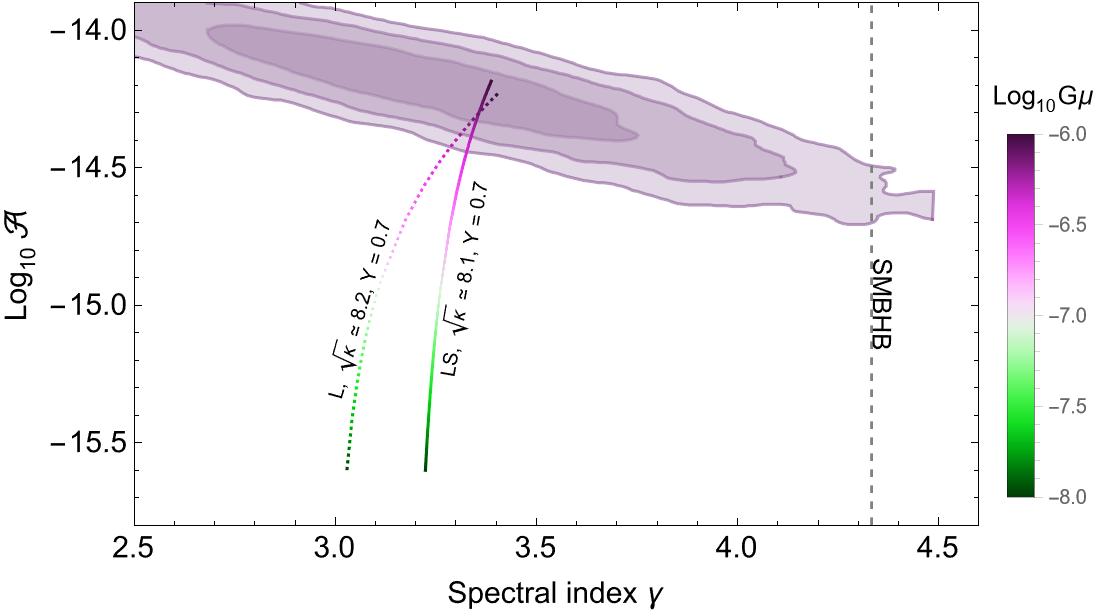}
\caption{GWs signal from metastable CSs with no unconfined magnetic flux (LS), confined magnetic flux (L), compared to the NANOGrav observations for $Y\simeq 0.7$, with $\sqrt{\kappa}\simeq 8.1$ (LS) and $8.2$ (L). Variation in $G\mu$ is shown by a vertical green-pink-toned color bar. The floral-violet shaded region shows the 3$\sigma$, 2$\sigma$ and 1$\sigma$ NANOGrav
posterior contours \cite{NANOGrav:2023gor}. The gray vertical line at $\gamma = 13/3$
represents the slope expected for the supermassive black hole binary (SMBHB) merger. Here $\gamma$ is the slope or the spectral index and $\mathcal{A}$ is the strain amplitude.}
\label{fig:AGamma}
\end{figure}

The decay rate per unit length of the CS is given by $\Gamma_d=(\mu/2\pi)\;\text{exp}(-\pi \kappa)$, where $\mu$ is the string tension, $\kappa=m_M^2/\mu$ is the metastability factor, and $m_M$ denotes the monopole mass \cite{Preskill:1992ck,Leblond:2009fq}. For $\sqrt{\kappa}\gtrsim 9$, the strings are practically stable. In Fig.~\ref{fig:GW} we present the GW spectra of the metastable CSs for $G\mu \approx 10^{-6}$ with $Y\simeq 0.3$ ($0.7$) and the metastability factor $\sqrt{\kappa}\simeq 8.0$ ($8.1$) for loops plus segments and $\sqrt{\kappa}\simeq 8.1$ ($8.2$) for loops only. In Fig.~\ref{fig:freqbins} the spectrum is shown with greater detail in the NANOGrav sensitivity band. Fig.~\ref{fig:AGamma} shows that the prediction is compatible with the NANOGrav signal at 1$\sigma$ level. For comparison, we have included the standard case of $Y=0$ in Fig.~\ref{fig:GW}, which one can see is ruled out by the LIGO O3 measurements. The allowed range to evade the LIGO O3 measurements is $Y\simeq 0.3-0.7$ for large loops corresponding to $\alpha=0.1$, as shown in Fig~\ref{fig:GW}.

%%%%%%%%%%%%%%%%%%%%%%%%%%%%%%%%%%%%%
%%%%%%%%%%%%%%%%%%%%%%%%%%%%%%%%%%%%%%%
%%%%%%%%%%%%%%%%%%%%%%%%%%%%%%%%%%%%%%%%%
%%%%%%%%%%%%%%%%%%%%%%%%%%%%%%%%%%%%%%%
%%%%%%%%%%%%%%%%%%%%%%%%%%%%%%%%

This range for the current is obtained using the fact that the two variables of the VOS model that describe the CS network evolution are the characteristic length $\mathbb{L}$, which is a measure of the energy density of the
network $\rho = \mu/\mathbb{L}^2$ ($\mu$ being the CS tension), and the
root-mean-squared (RMS) velocity $v$. For current-carrying CSs the characteristic length is defined in terms of the physical length $\mathbb{L}_\text{ph}=\mathbb{L}\,\sqrt{1-Y}$, and so the energy density $\rho=\mu/\mathbb{L}_\text{ph}^2$. The dimensionless GW energy density at higher frequencies is a flat plateau ($\Omega_{pl}$) which corresponds to loops formed in the radiation era and one finds \cite{Rybak:2022sbo}
\begin{align}\label{gwevol}
 \dfrac{\Omega^{CC}_\text{pl}}{\Omega^{S}_\text{pl}}=\dfrac{\dot{\rho}^{CC}_\text{pl}}{\dot{\rho}^{S}_\text{pl}}\sqrt{\dfrac{\Gamma^S\,\xi_r^{CC}}{\Gamma^{CC}\,\xi_r^S}}\;, 
\end{align}
where the superscripts $CC$ and $S$ represent the current-carrying and standard cases respectively. In Eq.~(\ref{gwevol}) $\xi_r$ is the average physical length and $\dot{\rho}$ is the energy density loss rate due to loop formation. These variables evolve with the current strength of the loops and affect the amplitude of the radiation plateau of the SGWB generated by the current-carrying CSs. The plateau decreases around $Y\simeq0.3-0.7$, and it then increases around $Y\gtrsim 0.8$. The evolution of $\xi_r$ shifts the spectrum towards higher frequencies (f) ($f \propto \xi_r$) compared to the standard case, as shown in Fig.~[8] of Ref.~\cite{Rybak:2022sbo}. This non-trivial evolution of the VOS variables for loops formed in the radiation era helps one to evade the LIGO O3 bounds and also provides a compelling explanation for the recent NANOGrav signal. Note that the contribution of segments is negligible at higher frequencies; for further details see Ref.~\cite{Buchmuller:2021mbb}. We have checked that following the above treatment, $G\mu$ values larger than $10^{-6}$ or so do not evade the LIGO O3 bounds.

\section{Conclusions}
\label{Con}
We have explored an $E_6$ based extension of the SM that yields two varieties of current-carrying metastable CSs, where the current in the two cases is associated either with the RHN zero modes or electrically charged particles. We find that in the presence of a suitable non-zero neutral current, the GW spectrum emitted by strings with a dimensionless string tension parameter $G \mu \sim 10^{-6}$ is compatible with the latest pulsar timing experiments as well as the LIGO O3 bounds. [In the absence of such current, the GW spectrum violates the LIGO O3 bounds unless $G \mu$ is lowered to $10^{-7}$ or so.]

%%%%%%%%%%%%%%%%%%%%%%%%%%%%%%%%%%%%%%%%
\section*{Acknowledgment}
AT is partially supported by the Bartol Research Institute, University of Delaware. QS thanks George Lazarides and Rinku Maji for helpful discussions.

%\clearpage
\bibliographystyle{JHEP}
\bibliography{bibliography.bib}

\end{document}